\documentclass[12pt]{article}
\begin{document}
\title{Interacting fermions and domain wall defects in $2+1$
  dimensions} \author{L.~Da~Rold~\footnote{Electronic address:
    darold@ib.cnea.gov.ar} C.~D.~Fosco~\footnote{Electronic address:
    fosco@cab.cnea.gov.ar} and A.~L{\'o}pez~\footnote{Electronic
    address: lopezana@cab.cnea.gov.ar}
  \\ \\
  {\normalsize\it $^a$Centro At\' omico Bariloche - Instituto Balseiro,}\\
  {\normalsize\it Comisi{\'o}n Nacional de Energ\'{\i}a At{\'o}mica}\\
  {\normalsize\it 8400 Bariloche, Argentina.}}  \date{\today}
\maketitle
\begin{abstract}
\noindent We consider a Dirac field in $2+1$ dimensions with a domain wall
like defect in its mass, minimally coupled to a dynamical Abelian
vector field.  The mass of the fermionic field is assumed to have just
one linear domain wall, which is externally fixed and unaffected by
the dynamics. We show that, under some general conditions on the
parameters, the localized zero modes predicted by the Callan and
Harvey mechanism are stable under the electromagnetic interaction of
the fermions.
\end{abstract}

\newpage
\section{Introduction}
It is a well-known fact that, in an odd~dimensional spacetime, a
domain wall defect in the mass term of a Dirac field induces a
fermionic zero mode localized on the defect~\cite{ch}.  This effect is
known to occur even in the presence of an external gauge field, if the
corresponding electromagnetic field is contained in the defect
hyperplane. Different aspects of this kind of system have been studied
both for static~\cite{p1,p2}, and dynamical~\cite{p3} defects. As far
as we know, however, possible effects due to interactions between the
fermions have not been considered for this system.  In this article,
we shall study the stability of this kind of configuration when the
electromagnetic interaction between the fermions is turned on.  That
the localization phenomenon should survive this interaction is not
{\em apriori\/} evident.  For example, for a static configuration, the
Coulomb repulsion between the localized charges could be so important
as to spread the charge density out over a large region, since the
charge density due to the zero mode shall induce an electromagnetic
field normal to the defect hypersurface.  On the other hand, we note
that our study may be thought of as a domain-wall analog of the
consideration of the self consistent vacuum currents in the presence
of vortices~\cite{li}. 

This paper is organized as follows: in section~\ref{model}, we introduce
the model and derive a self-consistent equation based in some approximations.
This equation is solved for two different mass profiles in  
section~\ref{examples}. Finally, in section~\ref{disc} we discuss the 
effects of the non-zero modes and present our conclusions.

\section{The model}\label{model}
The Euclidean action $S$, for the system we shall consider, is given
by
\begin{equation}
   \label{dfaction}
S \;=\; S_F \,+\, S_G
\end{equation}
where
\begin{equation}
   \label{defsf}
S_F\;=\; \int d^3x \; \bar{\psi}(x) [\not \!\partial
+ i e \not \!\!A(x) + M(x) ] \psi(x)
\end{equation}
is the fermionic action, and
\begin{equation}
   \label{defsg}
S_G\;=\; \int d^3x \; \frac{1}{4}  F_{\mu \nu} F_{\mu \nu} 
\end{equation}
the Maxwell action, which defines the gauge field dynamics.
\mbox{$x=(x_0,x_1,x_2)$} denote the Euclidean coordinates, and the
Hermitian $\gamma$ matrices are assumed to be in an irreducible $2 \times 2$
representation of the Dirac algebra, verifying the anticommutation
relations $\{\gamma_{\mu},\gamma_{\nu}\}= 2\,\delta_{\mu\nu}$.
The complete Green's functions can be derived from the generating
functional
$$
{\mathcal Z}[j_\mu;{\bar \eta},\eta ]\;=\;\int {\mathcal D}A_\mu
{\mathcal D}{\bar \psi} {\mathcal D}\psi \; \exp \{- S[{\bar\psi},
\psi; A]
$$
\begin{equation}
   \label{defz}
\,+\, \int d^3x [ j_\mu(x) A_\mu(x)\,+\, {\bar\eta}(x)\psi(x)
+ {\bar\psi}(x)\eta(x) \}
\end{equation}
where we included source terms for the gauge and fermionic fields.
The fermion mass is regarded as an external classical `field',
dependent on the $x_2$ coordinate only. We also fix the number of
defects to one, by requiring $M(x)$ to cross $0$ once, at $x_2=0$,
say.

By applying the property that the functional integral of a
(functional) derivative vanishes to equation (\ref{defz}), we derive
the `quantum equations of motion'
$$
0\;=\;\int {\mathcal D}A_\mu {\mathcal D}{\bar \psi} {\mathcal
  D}\psi \; \left[\frac{\delta S}{\delta A_\mu(x)}\; - j_\mu (x)
\right] \exp \{ - S[{\bar\psi},\psi; A_\mu]
$$
\begin{equation}
   \label{eq:eq1}
+ \int d^3x [j_\mu(x) A_\mu (x) + {\bar\eta}(x)\psi(x) +
{\bar\psi}(x)\eta(x)] \}
\end{equation}
for $A_\mu$, and
$$
0\;=\;\int {\mathcal D}A_\mu {\mathcal D}{\bar \psi} {\mathcal
  D}\psi \; \left[\frac{\delta S}{\delta {\bar \psi}(x)}\; - \eta(x)
\right] \exp \{ - S[{\bar\psi},\psi; A_\mu]
$$
\begin{equation}
   \label{eq:eq2}
+ \int d^3x [ j_\mu(x) A_\mu (x) + {\bar\eta}(x)\psi(x) +
{\bar\psi}(x)\eta(x) ] \} \;,
\end{equation}
for ${\bar\psi}$ (the adjoint equation is trivially obtained).  Taking
the functional derivative with respect to $\eta(y)$ in (\ref{eq:eq2}),
and putting all the external sources equal to zero afterwards, we find
that equations (\ref{eq:eq1}) and ({\ref{eq:eq2}) reduce to:
\begin{equation}
   \label{eq:eq3}
\partial_\mu F_{\mu\nu}(x) \;=\; J_\nu(x)
\end{equation}
and
\begin{equation}
   \label{eq:eq4}
\langle \, [\not\! \partial + i e \not \!\!A(x) + M(x)]
\psi(x) {\bar\psi}(y) \, \rangle \;=\; \delta (x-y) \;,
\end{equation}
where:
\begin{equation}
   \label{eq:defJ}
J_\nu(x)  \;=\;  i e \langle {\bar \psi}(x) \gamma_\nu \psi(x) \rangle
\end{equation}
and
\begin{equation}
   \label{eq:deff}
F_{\mu\nu}\;=\; \partial_\mu {\mathcal A}_\nu - \partial_\nu {\mathcal A}_\mu\;\;,\;\; {\mathcal A}_\mu
\,=\,\langle A_\mu\rangle \;.
\end{equation}
Equation (\ref{eq:eq3}) is an inhomogeneous `classical' Maxwell
equation, with the average gauge field \mbox{${\mathcal A}_\mu =
  \langle A_\mu \rangle$} playing the role of the classical gauge
field, and the average (vacuum) fermionic current $J_\mu$ as its
source.  Equation ({\ref{eq:eq4}) involves the expectation values
  \mbox{$\langle \psi {\bar\psi}\rangle$} and {$\langle A \psi
    {\bar\psi}\rangle$}. Of course, an exact treatment would require
  the use of an infinite set of coupled equations involving all the
  different Green's functions of the system.  In order to find a
  simpler and closed system of equations, we make the following
  approximation:
\begin{equation}
   \label{eq:approx}
\langle A_\mu (x) \psi(x) {\bar\psi}(y) \rangle \;\simeq\; \langle A_\mu (x)
\rangle \langle \psi (x) {\bar\psi} (y) \rangle \;=\;
{\mathcal A}_\mu(x) \, S_{\mathcal A}(x,y) \;,
\end{equation}
where we introduced $S_{\mathcal A}(x,y)$, which denotes the fermionic
propagator in the presence of an `external field' ${\mathcal A}(x)$,
which corresponds to the average gauge field.  This amounts to a sort
of mean field approximation, where the gauge field is treated
classically. To make the approximation involved more explicit, we note
that the (exact) three point function appearing in (\ref{eq:approx})
can be written in the equivalent form:
\begin{equation}
   \label{eq:approx1}
\langle A_\mu (x) \psi(x) {\bar\psi}(y) \rangle \;=\; \int {\mathcal D}A
\,A_\mu (x)\, \langle x |( \not \! \partial + i e \not \!\! A + M )^{-1}| y \rangle
\, e^{- S_G[A] - \Gamma_F [A]}
\end{equation}
where
\begin{equation}
   \label{eq:defgf}
\Gamma_F [A] \;=\; - {\rm log}\det[\not\!\partial + i e \not \!\! A + M ]\;.
\end{equation}
The approximation (\ref{eq:approx}) is obtained from
(\ref{eq:approx1}) by replacing $A$ by its saddle point value. Namely,
the approximation amounts to using the (leading) saddle point
approximation, where the `action' which is minimized at the saddle
point is the bare Maxwell action plus an effective contribution
$\Gamma_F[A]$ coming from the fermionic determinant.

Equation (\ref{eq:approx}) is sufficient to close the system of
equations, since then (\ref{eq:eq4}) becomes:
\begin{equation}
   \label{eq:eq4ap}
[\not\! \partial + i e \not\!\!{\mathcal A}(x) + M(x)] S_{{\mathcal A}}\;=\; \delta (x-y) \;.
\end{equation}
It is now important to realize that the average current can be
expressed as a functional of ${\mathcal A}$, as follows:
\begin{equation}
   \label{eq:eq5}
J_\mu (x) \;=\;i e\; {\rm tr}\left[\gamma_\mu \langle \psi(x){\bar\psi}(x)\rangle \right] \;=\;
- i e\; {\rm tr} \left[\gamma_\mu S_{\mathcal A}(x,x) \right] \;.
\end{equation}
Equation (\ref{eq:eq3}), together with (\ref{eq:eq5}), define a closed
system of equations, which allows us to find the average gauge field
${\mathcal A}$, and then the current density induced in that
background.  The equation that determines ${\mathcal A}$ is obtained
by replacing $J_\mu$ by its expression (\ref{eq:eq5}) in
(\ref{eq:eq3}):
\begin{equation}
   \label{eq:eqa}
\partial_\mu F_{\mu\nu}(x)\;=\;-i e\; {\rm tr} \left[\gamma_\mu S_{\mathcal A}(x,x)\right] \;,
\end{equation}
which, in general, and depending on the approximation used to evaluate
$S_{\mathcal A}$, will be a non-linear integro-differential equation.
The non-linearity comes from the fermionic propagator $S_{\mathcal
  A}$, which is defined as :
\begin{equation}
S_{\alpha\beta}(x,y)=\langle x,\alpha \mid {\mathcal D}^{-1} \mid y, \beta \rangle ,
\end{equation}
where ${\mathcal D}=(\not \! \partial + i e \not\!\! {\mathcal A} + M
)$.

We shall now look for particular solutions of the coupled set of
equations, under some restrictions and simplifying approximations.  We
shall restrict ourselves to {\em static\/}, purely electric solutions,
with no electric current (hence, no magnetic field). In the Coulomb
gauge, the only remaining component for the (average) gauge field is
${\mathcal A}_0$, which is determined by the equation
\begin{equation}
   \label{eq:gauss}
\nabla^2 V \;=\; - i e\; tr[\gamma_0 S_V(x,x)] \;,
\end{equation}
where $V\;=\;{\mathcal A}_0$.

Our approach to solve the system of equations shall be to first
evaluate the fermionic propagator in the external potential $V$. Then,
we shall find the corresponding vacuum charge density as a functional
of $V$, and insert it into the Gauss law (\ref{eq:gauss}) to determine
$V$. The resulting $V$ can then be used to fix the precise form of the
charge density. We will be able to say that there are localized modes
if the system admits solutions where the charge density is confined to
a small region around the defect. Of course, we shall have to make
some assumptions also on the allowed boundary conditions for the
fields. The choice of these conditions is also part of the kind of
ansatz used, and also on the amount of generality one wants to
introduce into the treatment.

To find the fermion propagator in the presence of the external field
$V$, we shall use the perturbative expansion of ${\mathcal D}^{-1}$ in
powers of $V$, namely, we decompose ${\mathcal D}$ as follows:
\begin{equation}
{\mathcal D} = {\mathcal D}_0 + {\mathcal V},
\end{equation}
where
\begin{equation}
{\mathcal D}_0 = \not\!\partial + M(x)
\end{equation}
and
\begin{equation}
{\mathcal V} \;=\; i e \gamma_0  V(x)\;.
\end{equation}
Thus, ${\mathcal D}^{-1}$ is naturally expanded as:
\begin{equation}
{\mathcal D}^{-1} = {\mathcal D}_0 ^{-1} - {\mathcal D}_0 ^{-1}
{\mathcal V} {\mathcal D}_0 ^{-1} + {\mathcal D}_0 ^{-1} {\mathcal V}
{\mathcal D}_0 ^{-1} {\mathcal V} {\mathcal D}_0 ^{-1}
-...\label{perturbaciones}
\end{equation}
We note that the `free' propagator ${\mathcal D}_0^{-1}$ includes the
mass field and its space dependence exactly. This must be so, since
the defect changes the spectrum of the Dirac field, an effect that
cannot be described perturbatively. To find the inverse of ${\mathcal
  D}_0$, we use the equivalent expression:
\begin{equation}
{\mathcal D}_0 ^{-1}=({\mathcal D}_0 ^{\dagger}{\mathcal D}_0)
^{-1}{\mathcal D}_0 ^{\dagger} \label{SdeH}.
\end{equation}
which requires finding the inverse of the Hermitian operator
\begin{equation}
   \label{eq:defh0}
{\mathcal H}_0 \;=\; {\mathcal D}_0^\dagger {\mathcal D}_0 \,.
\end{equation}
This is a much simpler task than inverting ${\mathcal D}_0$, and it
allows one to dimensionally reduce the problem. To see this, we follow
the procedure of \cite{p1}, of which we give a lightning review here.
First we write:
\begin{equation}
{\mathcal D}_0 = (a + \widehat{/\!\!\!\partial} ) P_L + (a^{\dagger} + \widehat{/\!\!\!\partial}
 ) P_R \label{Ddea} ,
\end{equation}
where $\widehat{\not\!\partial} = \gamma_0 \partial_0 + \gamma_1
\partial_1$.  We define the operators $a^{\dagger}$ and $a$, that act
on functions of the $x_2$ coordinate as
\begin{equation}
   \label{op2}
a = \partial_2 + M \;\;\;\; a^{\dagger} = -\partial_2 + M,
\end{equation}
and the projectors $P_L$, $P_R$:
\begin{equation}\label{op3}
P_L = \frac{1 + \gamma_2}{2}, \;\;\;\;\;\;\;\;\;\; P_R = \frac{1 -
\gamma_2}{2}.
\end{equation}
These projectors behave like chirality projectors from the point of
view of the $1+1$ dimensional theory which describes the chiral zero
mode.  This decomposition makes it possible to disentangle the
dynamics corresponding to the $x_2$ coordinate from the coordinates
$\hat x = ( x_0 , x_1 )$.  The `dimensional reduction' can be seen to
arise at the level of the operator ${\mathcal H}_0$:
\begin{equation}
   \label{Hdea}
{\mathcal H}_0 =
(h - \widehat{\not\!\partial}^2) P_L + (\tilde{h} - \widehat{\not\!\partial}^2) P_R \;,
\end{equation}
where
\begin{equation}
h \;=\; a^{\dagger} a \;\;\;\;\;\; \tilde{h} \;=\; a\,a^{\dagger}\;.
\end{equation}
To expand the fermionic fields, we define $\phi_n$ and
$\widetilde{\phi}_n$, eigenstates of the operators $h$ and
$\tilde{h}$, respectively. We denote by $\lambda_n^2$ their (common)
eigenvalues:
\begin{equation}
h \phi_n = \lambda_n^2 \phi_n , \;\;\;\;\;\;
\tilde{h} \widetilde{\phi}_n = \lambda_n^2 \widetilde{\phi}_n ,
\end{equation}
\begin{equation}
\langle \phi_n \mid \phi_m \rangle = \delta_{nm}, \;\;\;\;\;\;\;
\langle \widetilde{\phi}_n \mid \widetilde{\phi}_m \rangle = \delta_{nm} ,
\end{equation}
since the spectra coincide, except for $\lambda_n = 0$, and the
eigenvalues are of course positive.  The $\lambda_n = 0$ eigenvalue
will, by assumption, be present only for $h$.  This will depend of
course on the mass profile near the defect, i.e. the zero of the mass.
Since the sign of $\lambda_n$ is arbitrary, we take it positive by
convention.

Thus, the fermionic fields can be expanded as:
\begin{equation}
\psi (\hat x , x_2 ) = \sum_n [\phi_n(x_2) \psi_L^{(n)}(\hat x) +
\widetilde{\phi}_n(x_2) \psi_R^{(n)}(\hat x)] ,
\end{equation}
\begin{equation}
\overline{\psi} (\hat x , x_2 ) = \sum_n [\overline{\psi}_L^{(n)}(\hat x) \phi_n^{\dagger}(x_2)
+ \overline{\psi}_R^{(n)}(\hat x) \widetilde{\phi}_n^{\dagger}(x_2)] .
\end{equation}
The spinors that carry the dependence on $\hat x$ are defined by:
\begin{equation}
\psi_{L,R}^{(n)}(\hat x)\,=\,P_{L,R} \psi^{(n)}(\hat x)\;\;,\;\;\;\;\;
\overline{\psi}_{L,R}^{(n)}(\hat x)\,=\,\overline{\psi}^{(n)}(\hat x) P_{R,L}\;,
\end{equation}
where $\psi_{L,R}^{(n)}$ denotes a general bidimensional fermionic
field (one for each value of the index $n$). In terms of this
expansion, the fermionic action becomes:
\begin{equation}
   \label{eq:actexp}
S \;=\; S_L^{(0)} \,+\, \sum_n  S^{(n)}
\end{equation}
where $S_L^{(0)}$ denotes the action for a chiral left-handed fermion
in $1+1$ dimensions, while $S^{(n)}$ is a massive Dirac action, also
in $1+1$ dimensions, with a mass equal to $\lambda_n$ (the sign of the
mass is irrelevant in $1+1$ dimensions).

Since
\begin{equation}
{\mathcal H}_0 ^{-1}=
(h - \widehat{/\!\!\!\partial}^2)^{-1} P_L + (\tilde{h} - \widehat{/\!\!\!\partial}^2)^{-1} P_R ,
\end{equation}
the free propagator becomes
\begin{equation}
   \label{Ddeop}
{\mathcal D}_0 ^{-1}= (h - \widehat{/\!\!\!\partial}^2)^{-1} P_L
(a^{\dagger} - \widehat{/\!\!\!\partial}) + (\tilde{h} -
\widehat{/\!\!\!\partial}^2)^{-1} P_R (a - \widehat{/\!\!\!\partial}) \;.
\end{equation}
Translation invariance along the $x_0$ and $x_1$ coordinates suggests
the use of a potential depending only on $x_2$, ${\mathcal
  V}\,=\,{\mathcal V}(x_2)$.  To find the propagator in configuration
space, we need to evaluate the following expression:
\begin{equation}
S_{\alpha\beta}(x,y)=({\mathcal D}_0 ^{-1})_{\alpha\beta}(x,y)-({\mathcal D}_0 ^{-1}{\mathcal V}
{\mathcal D}_0 ^{-1})_{\alpha\beta}(x,y) + ...
\end{equation}
with ${\mathcal V}={\mathcal V}(x_2)$. In the perturbative expansion
for the propagator, we insert expansions of the identity constructed
with intermediate states corresponding to eigenstates of the operator
${\mathcal H}_0$. Using the fact that each eigenvalue $\lambda_n$
corresponds to the effective mass of a two dimensional mode, and that
the lowest mode is massless (the zero mode), it is natural to keep
only the zero mode in the intermediate states as a first
approximation. Note that the mass $\lambda_n$ of the non zero modes is
separated from the zero mode by a finite gap whose magnitude is
controlled by the profile of the mass near the defect (see ref.
\cite{p1}). With this in mind, we shall first use the leading
approximation of keeping just the zero mode, and then make a
quantitative evaluation of the error involved in this procedure, by
including the correction corresponding to the lowest massive mode. On
the other hand, we shall keep the full dependence in the potential,
namely, we shall use no truncation for the perturbative series in
${\mathcal V}$.  To implement this approximation, we introduce
projectors $P_0$ along the zero mode. They are explicitly given by
\begin{equation}
P_0 = \phi_0 \phi_0^{\dagger} \sum_n \psi_L^{(n)} \overline{\psi}_L^{(n)}\,.
\end{equation}
Taking this into account, after some algebra one can show that, in
this approximation, the propagator is given by:
\begin{equation}
S_{\alpha\beta}(x,y) \simeq \phi^{\dagger}_0(x_2)\phi_0(y_2)   \langle x_0,x_1,\alpha \mid \frac{\widehat{/\!\!\!\partial} +
ie\;\gamma_0 V_{0,0} P_L}{(\widehat{\partial} + ie\;\gamma_0 V_{0,0})^2}\mid y_0,y_1, \beta \rangle\,.
\end{equation}
In this expression there appears the average of $V$ in the zero mode
which is denoted by:
\begin{equation}
V_{0,0}\;=\;\langle \phi_0 | V | \phi_0 \rangle\;.
\end{equation}

It is worth noting that this result is approximate in the sense that
only the zero mode has been included, but all the powers of $A_\mu$
have been added, as it is evident from the non-linear dependence of
the propagator on $A_\mu$.  The charge density is evaluated by
multiplying by $\gamma_0$, taking the Dirac trace, and finally
calculating the coincidence limit $x \to y$. Inserting the result so
obtained for the charge density as a functional of the potential into
(\ref{eq:gauss}) yields:
\begin{equation}
   \label{ecdif}
\frac{\partial^2}{\partial x_2^2}V(x_2)\;=\; \phi_0(x_2)\phi^{\dagger}_0(x_2) \int \frac{d^2k}{2\pi}
\frac{-i\;k_0+ie\;V_{0,0}}{(-i\;k_0+ie \gamma_0 V_{0,0})^2}\;.
\end{equation}
The momentum space integral has both linear and logarithmic
divergences. Using a symmetric limit kind of regularization, we see
that (\ref{ecdif}) can be expressed as:
\begin{equation}
   \label{ecdif1}
\frac{\partial^2}{\partial x_2^2}V(x_2) =
\phi_0(x_2)\phi^{\dagger}_0(x_2)\frac{e^2\;V_{0,0}}{2}\;.
\end{equation}
It is remarkable that, as a consequence of the fact that we are only
keeping the zero mode, the expression for the charge density becomes
linear in the potential. This happens in spite of the fact that we
have kept all the powers of the potential in the fermionic propagator,
since the result is a consequence of the fact that massless two
dimensional $QED$ is exactly solvable~\cite{sch}, with the exact
fermionic determinant being quadratic in the gauge field.

We have obtained an integro-differential equation involving
derivatives of $V$ and its average on the lowest energy mode. To solve
it self-consistently, we first derive from (\ref{ecdif1}) (by
integration) an equation for $V$, depending also the average of the
potential. Then, as a second step, we shall insert this average into
(\ref{ecdif1}) in order to obtain the explicit profile of the
potential as a function of $x_2$.  At this point, it is clear that the
existence of a self-consistent solution depends on the particular form
of the zero modes appearing in eq.  (\ref{ecdif1}). This differential
equation will have a solution only if the charge density is localized
in such a way that the integrals involved are well defined. In
particular, the zero modes need to be localized around the defect.  It
was shown in reference~\cite{ch} that in $2n+1$ dimensions the zero
mode has the form
\begin{equation}
\eta\;e^{-\int_a^{x_2}dy\;M(y)},\label{kallan}
\end{equation}
where $\eta$ is an spinor independent of $x_2$.
\section{Examples}\label{examples}
In what follows we will discuss the possible solutions of eq.
(\ref{ecdif1}) for two different kinds of mass profiles.
\begin{itemize}
\item Step-like defect.
 
 Given a mass of the form:
\begin{equation}
M(x_2)\;=\;\Lambda (2\Theta (x_2)-1)\,.
\end{equation}
where $\Lambda$ is a constant with the dimensions of a mass, and
$\Theta$ is the Heaviside function, there is only one zero
mode~\cite{p1}, which can be explicitly written as:
\begin{equation}
\phi_0(x_2)=\Lambda^{\frac{1}{2}}\;e^{-\Lambda|x_2|}\;.
\end{equation}
In this case, the differential equation becomes
\begin{equation}
   \label{eq:g1}
\frac{\partial^2}{\partial x_2^2}V(x_2)\;=\;\frac{1}{2} \, \Lambda
e^2 V_{0,0}\; e^{-2\Lambda|x_2|}\;,
\end{equation}
and integrating it twice we obtain for the potential $V$
\begin{equation}
\label{vx2}
V(x_2)=a+\frac{1}{8 \Lambda}\, e^2 V_{0,0}\; e^{-2\Lambda|x_2|}\;,
\end{equation}
where $a$ is a constant, to be related later to the chemical
potential.  In this expression we have not included a term that
corresponds to a constant electric field in the $x_2$ direction,
because it could be eliminated by choosing appropriate boundary
conditions (such as vanishing density of charges at infinity).

In order to find a self-consistent solution for the potential we
evaluate the expectation value of $V$, which is expressed by
(\ref{vx2}), in the zero mode
\begin{equation}
V_{0,0}\;=\;a\,+\, \frac{1}{8 \Lambda}\, e^2 V_{0,0}\int_{-\infty}^{\infty} dx
|\phi_0(x)|^2 \;e^{-2\Lambda|x|}\;.
\end{equation}
Thus, $V_{0,0}$ is easily seen to be given by
\begin{equation}
   \label{eq:g2}
V_{0,0}\;=\;\frac{a}{1-\frac{e^2}{16\Lambda}}\;.
\end{equation}
Therefore, the potential written in terms of the zero modes results
\begin{equation}
V({x_2})=|\phi_0(x)|^2{\frac{2ae^2}{16\Lambda^2-\Lambda e^2}}.
\end{equation}
Notice that the solution is only stable if the electromagnetic
coupling constant and the mass coupling constant satisfy the bound:
$e^2<16\Lambda$, which means that the strength of the interaction
(repulsion) between the electrons cannot be larger than the scale
given by the height of the defect. We note that `stability' refers
here to the property of having a confining potential.  We see that in
this case, i.e., for an step-like mass and keeping only the zero
energy mode, there exist a self-consistent solution for the fermionic
interaction potential. In other words, even in the case of interacting
electrons, the fermions are localized in the $x_2$ direction and can
only move along the defect.

The interpretation of $a$ as a chemical potential proceeds from the
fact that the Gauss law (\ref{eq:g1}), combined with (\ref{eq:g2}),
means that the charge density of the configuration is
\begin{equation}
\rho (x_2)\;=\;  a \, \frac{\Lambda e^2}{2(1-\frac{e^2}{16\Lambda})}
e^{-2 \Lambda |x_2|}\;,
\end{equation}
and (by integrating over $x_2$) one sees that the total charge is
proportional to the constant $a$.

\item Linear defect.
  
  Assuming than the mass can be expanded as a power series in $x_2$,
  for small enough $x_2$ we only keep the first order term:
\begin{equation}
M(x_2)=M'(0)\;x_2,
\end{equation}
where we assume $M'(0)\neq 0$ being $M'$ the first derivative of the
mass.  For this mass profile we can still find the zero mode by
defining~\cite{p1}
\begin{equation}
h\;=\;-\partial_2^2\,-\, M'\,+\,M^2 x_2^2\;.
\end{equation}
which is an harmonic oscillator Hamiltonian. The lowest energy mode
is:
\begin{equation}
\phi_0(x_2)=(\frac{|M'|}{\pi})^{1/4}e^{-\frac{|M'|}{2}x_2^2}.
\end{equation}
Following the same steps as in the previous example we find that the
potential can be written in terms of the zero mode as
\begin{equation}
V(x_2)=a+\int_B^{x_2} dy \int_A^y dz |\phi_0(z)|^2(\frac{e^2a}{2-e^2C}),
\end{equation}
where
\begin{equation}
C=\int_{-\infty}^{\infty}dx_2 |\phi_0(x_2)|^2 \int_B^{x_2} dy \int_A^y dz |\phi_0(z)|^2.
\end{equation}
Thus we see that also in this case there exists a self-consistent
solution for the Gauss law, for a charge density localized around the
defect.  However there is a necessary condition for the existence of
this localized mode. The wave function of the zero mode has to vanish
rapidly outside the region of the space where the mass can be
approximated linearly. A quantitative criterion for the validity of
this condition can be found in reference \cite{p1}.

\end{itemize}

In summary, up to know we have shown the existence of localized
solutions if we keep only the lowest energy modes in the expansion of
the fermionic propagator. This solution depends on the mass profile,
and it is non-perturbative in the electromagnetic interaction between
the fermions.  We have neglected the (more energetic) massive modes
based on the fact that the terms on the action that come from these
modes go as $\frac{1}{\lambda_n^2}$, where $\lambda_n$ is the mass of
the mode~\cite{p1}. Therefore, for a large and steep enough mass, our
approximation will be valid. In particular, for a linear defect, the
mass of the modes is proportional to the slope of the mass profile.
Therefore, by changing this slope we could make $\lambda_n$
arbitrarily large.

\section{Effect of the massive modes}\label{disc}
We shall now study the problem of including one massive mode in our
calculation in order to check whether there still exist localized
solutions or not.  It will provide also a quantitative idea about the
error involved in considering only the lowest energy modes.

We proceed as follows: in the perturbative expansion for the fermion
propagator (\ref{perturbaciones}), we consider only the projection of
the operators ${\mathcal D}_0^{-1}$ and ${\mathcal V}$ onto the two
lowest energy modes. Then each factor ${\mathcal D}_0^{-1}$
contributes with
 \begin{equation}
{\mathcal D}_0^{-1} \simeq \widetilde{\phi}_1\phi^{\dagger}_1
\lambda_1 P_L - \widetilde{\phi}_1 \widetilde{\phi}^{\dagger}_1
\widehat{/\!\!\!\partial} P_R + \phi_1
\widetilde{\phi}^{\dagger}_1 \lambda_1 P_R - \phi_1
\phi^{\dagger}_1 \widehat{/\!\!\!\partial} P_L -
\phi_0\phi^{\dagger}_0 \widehat{/\!\!\!\partial} P_L
\label{ordencero}.
\end{equation}
For an even $V(x_2)$, selection rules imply the vanishing of the
matrix elements $V_{n,m}$, $V_{n,\tilde m}$ and $V_{\tilde n,\tilde
  m}$.

Replacing (\ref{ordencero}) into (~\ref{perturbaciones}), and keeping
only the non-vanishing matrix elements of $V$, the first order
correction in ${\mathcal V}$ to the fermion propagator (i.e.,
correction to (~\ref{ordencero})) is:
$$
{\mathcal D}_0 ^{-1}{\mathcal V}{\mathcal D}_0 ^{-1} \simeq
\frac{{ie\widetilde{\phi}_1 \phi^{\dagger}_1}}{{(\lambda_1^2 -
    {\widehat\partial}^2)^2}} [-\widehat{/\!\!\!\partial} \lambda_1
\gamma^0 V^0_{\tilde1,\tilde1} - \lambda_1 \gamma^0 V_{1,1}
\widehat{/\!\!\!\partial}]P_L$$
$$+\frac{{ie\widetilde{\phi}_1\phi^{\dagger}_0}}{{(\lambda_1^2-\widehat\partial^2)\widehat\partial^2}}
[-\lambda_1 \gamma^0V_{1,0} \widehat{/\!\!\!\partial}]P_L
+\frac{ie\widetilde{\phi}_1\widetilde{\phi}^{\dagger}_1}{(\lambda_1^2-\widehat\partial^2)^2}
[\lambda_1^2
\gamma^0V_{1,1}+\widehat{/\!\!\!\partial}\gamma^0V_{\tilde1,\tilde1}\widehat{/\!\!\!\partial}]P_R
$$
$$+\frac{ie\phi_1\widetilde{\phi}^{\dagger}_1}{(\lambda_1^2-\widehat
  \partial^2)^2} [-\widehat{/\!\!\!\partial} \lambda_1
\gamma^0V_{1,1}- \lambda_1 \gamma^0V_{\tilde1,\tilde1}
\widehat{/\!\!\!\partial}]P_R +
\frac{ie\phi_0\widetilde{\phi}_1^{\dagger}}{(\lambda_1^2-\widehat
  \partial^2)\widehat\partial^2} [-\widehat{/\!\!\!\partial}\lambda_1
\gamma^0V_{1,1} ]P_R$$
$$+\frac{ie\phi_1 \phi^{\dagger}_1}{(\lambda_1^2-\widehat
  \partial^2)^2} [\lambda_1^2
\gamma^0V_{\tilde1,\tilde1}+\widehat{/\!\!\!\partial}\gamma^0V_{1,1}\widehat{/\!\!\!\partial}]P_L
+ \frac{ie\phi_1 \phi^{\dagger}_0}{(\lambda_1^2-\widehat
  \partial^2)\widehat \partial^2}
[\widehat{/\!\!\!\partial}\gamma^0V_{1,0}\widehat{/\!\!\!\partial}]P_L$$
\begin{equation}
+ \frac{i e \phi_0\phi_1^{\dagger}}{(\lambda_1^2-\widehat
\partial^2)\widehat \partial^2} [\widehat{/\!\!\!\partial}\gamma^0V_{0,1}\widehat{/\!\!\!\partial}]P_L
+ \frac{i e \phi_0\phi_0^{\dagger}}{(\widehat
\partial^2)^2}  [\widehat{/\!\!\!\partial}\gamma^0V_{0,0}\widehat{/\!\!\!\partial}]P_L.
\end{equation}
Notice that in this case it is not possible to obtain a
non-perturbative expression for the fermion propagator due to the fact
that we are taking into account massive modes as well as the massless
one. In order to write the Gauss law we need to compute
$$
tr(\gamma_0{\mathcal D}^{-1}) \simeq -\frac{ie(\widetilde{\phi}_1
  \widetilde{\phi}^{\dagger}_1+\phi_1
  \phi^{\dagger}_1)}{(\lambda_1^2-\widehat \partial^2)} \partial_0 +
\frac{ie\phi_0\phi_0^{\dagger}}{(\widehat \partial^2)} \partial_0
$$
$$+ \frac{ie\widetilde{\phi}_1
  \widetilde{\phi}^{\dagger}_1}{(\lambda_1^2-\widehat \partial^2)^2}
[\lambda_1^2 V_{1,1}+(2\partial_0^2-{\widehat
  \partial}^2)V_{\tilde1,\tilde1}]+
\frac{ie\phi_1 \phi^{\dagger}_1}{(\lambda_1^2-\widehat\partial^2)^2}
[\lambda_1^2V_{\tilde1,\tilde1}+(2\partial_0^2-{\widehat
  \partial}^2)V_{1,1}]
$$
\begin{equation}
+ \frac{ie(\phi_1 \phi^{\dagger}_0 + \phi_0 \phi^{\dagger}_1)}
{(\lambda_1^2-{\widehat\partial}^2){\widehat\partial^2}}
[(2\partial_0^2-{\widehat\partial}^2)V_{0,1}] +
\frac{ie\phi_0\phi_0^{\dagger}}{({\widehat\partial}^2)^2}
[(2\partial_0^2-{\widehat
\partial}^2)V_{0,0}]. \label{trmasiva}
\end{equation}
Taking the Fourier transform in the above expression and regularizing
the integrals by a symmetric limit, the Gauss law becomes
\begin{equation}
\frac{\partial^2}{\partial x_2^2}V(x_2)
=\widetilde{\phi}_1(x_2)\widetilde{\phi}^{\dagger}_1(x_2)\frac{e^2\;V_{1,1}}{2}+
\phi_1(x_2)\phi^{\dagger}_1(x_2)\frac{e^2\;V_{\tilde1,\tilde1}}{2}.
\end{equation}
Thus we have obtained a differential equation whose solution will
depend on the localization properties of the fermionic modes around
the defect.

In the case of a mass that can be approximated by a linear function of
$x_2$ near the defect, it is simple to check that~\cite{p1}
\begin{equation}
\phi_n=\widetilde{\phi}_{n+1},
\end{equation}
therefore there is only one zero mode, and the Gauss equation becomes
\begin{equation}
\frac{\partial^2}{\partial x_2^2}V(x_2)
=\phi_0(x_2)\phi^{\dagger}_0(x_2)\frac{e^2\;V_{1,1}}{2}+
\phi_1(x_2)\phi^{\dagger}_1(x_2)\frac{e^2\;V_{0,0}}{2}.
\end{equation}
Integrating this expression we find
\begin{equation}
V(x_2)=a+(\frac{e^2V_{1,1}}{2})\int_B^{x_2} dy \int_A^y dz
|\phi_0(z)|^2+(\frac{e^2V_{0,0}}{2})\int_B^{x_2} dy \int_A^y dz
|\phi_1(z)|^2.
\end{equation}
Once again, we look for the self-consistent solutions for the
expectation values of the potential. When computed on the two lowest
energy modes, they are given by the solution of the equations:
\begin{equation}
V_{i,i}=a+(\frac{e^2V_{1,1}}{2})D_{i0}+(\frac{e^2V_{0,0}}{2})D_{i1},
\label{vij}
\end{equation}
where $i=0,1$ and $D_{ij}$ are:
\begin{equation}
2D_{ij}=\int_{-\infty}^{\infty} dx_2 |\phi_i(x_2)|^2\int_B^{x_2} dy
\int_A^y dz |\phi_j(z)|^2.
\end{equation}
Solving (\ref{vij}) we obtain
\begin{equation}
V_{0,0}=a\frac{1-e^2D_{10}+e^2D_{00}}
{(1-e^2D_{01})(1-e^2D_{10})-e^4D_{11}D_{00}}, \nonumber
\end{equation}
\begin{equation}
V_{1,1}=a\frac{1-e^2D_{01}+e^2D_{11}}
{(1-e^2D_{01})(1-e^2D_{10})-e^4D_{11}D_{00}}.
\end{equation}
We have found that, in the case of a linear mass, there exist a
self-consistent solution of the Gauss equation to first order in the
interaction potential, if we include apart from the zero mode, one
massive mode.  Notice that, for a linear mass around the defect,
$\phi_n$ and $\widetilde{\phi}_n$ are harmonic oscillator eigenstates.
Far enough from the defect, the eigenstates decay exponentially (as a
Gaussian function), ensuring that the charge density is localized
around the defect in such a way that there is a solution for the Gauss
equation.  Obviously all the caveats regarding the range of validity
of approximating the mass by a linear function, that we mention in the
previous case, must be taken into account here.

Summarizing, we have considered a Dirac field in $2+1$ dimensions with
a domain wall like defect in its mass, minimally coupled to a
dynamical Abelian vector field.  The mass of the fermionic field is
assumed to have just one linear domain wall, externally fixed and
unaffected by the dynamics.  In the absence of electromagnetic
interactions among the fermions, it is a well known fact that
localized zero modes exist on the defect~\cite{ch}. We have studied
here the effect of the fermionic interactions on these modes showing
that, under some general conditions on the parameters, the localized
zero modes stable under the electromagnetic interactions of the
fermions.

\section*{Acknowledgments}

This work is partially supported by CONICET (Argentina), by ANPCyT
through grant No.$03-03924$ (AL), and by Fundaci{\'o}n Antorchas
(Argentina).

\newpage

\end{document}